\documentclass[12pt,letterpaper]{article}
\pdfoutput=1
\usepackage{jheppub}
\usepackage{amsfonts, amsthm}
\usepackage[english]{babel}
\usepackage[utf8]{inputenc}
\usepackage{slashed}
\usepackage{mathrsfs}
\usepackage{amssymb}
\usepackage{color}
\hypersetup{unicode}

\newcommand{\eq}{\begin{equation}}
\newcommand{\feq}{\end{equation}}
\newcommand{\eqn}{\begin{eqnarray}}
\newcommand{\feqn}{\end{eqnarray}}

\newcommand{\ma}[1]{\mbox{$\mathcal{#1}$}}

\newcommand{\D}{{\rm d}}

\title{First order flow equations for nonextremal black holes in AdS (super)gravity}

\author{Dietmar Klemm and Marco Rabbiosi}
\affiliation{Dipartimento di Fisica, Universit\`a di Milano, Via Celoria 16, 20133 Milano, Italy, \\
and \\
INFN, Sezione di Milano, Via Celoria 16, 20133 Milano, Italy.
}
\emailAdd{dietmar.klemm@mi.infn.it}
\emailAdd{marco.rabbiosi@mi.infn.it}
\preprint{IFUM-1054-FT}

\abstract{We consider electrically charged static nonextremal black holes in $d$-dim\-ens\-ional
Einstein-Maxwell-(A)dS gravity, whose horizon is a generic Einstein space in $d-2$ dimensions.
It is shown that
for this system the Hamilton-Jacobi equation is exactly solvable and admits two branches of solutions.
One of them exhibits a non-simply connected domain of integration constants and does not reduce to
the well-known solution for the $d=4$ BPS case. The principal functions generate two first order flows
that are analytically different, but support the same general solution. One of the two sets of flow
equations corresponds to those found by L\"u, Pope and V\'azquez-Poritz in hep-th/0307001 and (for 
$d=4$ and $\Lambda=0$) by Miller, Schalm and Weinberg in hep-th/0612308. This clarifies also the
reason for the very existence of first order equations for nonextremal black holes, namely, they are just
the expressions for the conjugate momenta in terms of derivatives of the principal function in
a Hamilton-Jacobi formalism. In the last part of our paper we analyze how much of these integrability 
properties generalizes to matter-coupled $N=2$, $d=4$ gauged supergravity.}


\begin{document}
\maketitle
\flushbottom

\section{Introduction}

Exact solutions to Einstein's field equations and their supergravity generalizations have been playing,
and continue to play, a crucial role in many important developments in general relativity, black hole
physics, integrable systems, string theory and quantum gravity. Being highly nonlinear, coupled
partial differential equations, these are notoriously difficult to solve, sometimes even in presence
of a high degree of symmetry, for instance in supergravity where one has typically many other fields
in addition to the metric. If one imposes that some fraction of supersymmetry be preserved, the
construction of solutions simplifies considerably, since one has to solve only the
first order Killing spinor equations instead of the full higher order equations of motion.

In the last years however it was shown (see \cite{Lu:2003iv,Miller:2006ay,Ceresole:2007wx,
Andrianopoli:2007gt,LopesCardoso:2007qid,Janssen:2007rc,
Cardoso:2008gm,Perz:2008kh,Ceresole:2009iy,
Galli:2011fq,Barisch:2011ui,Klemm:2012vm,
Gnecchi:2012kb,Gnecchi:2014cqa,Cardoso:2015wcf,Lindgren:2015lia, Klemm:2016wng} for an (incomplete) list of
references) that sometimes also non-BPS- and even nonextremal black holes satisfy certain first order 
equations, which typically arise by writing the potential of a one-dimensional mechanical system (to which 
the supergravity action boils down if one has enough symmetry)  in terms of a
`superpotential'\footnote{An analysis of nonextremal solutions was used in \cite{Gursoy:2008za} to study
Einstein-dilaton black holes. In the most general case with a vector and scalar this was done
in \cite{Kiritsis:2012ma}, where the motivation was the computation of the finite temperature/finite
density effective potential in holography. Recently the zero temperature case was analyzed in full generality
in Einstein-dilaton gravity \cite{Kiritsis:2016kog} in order to find the most general RG flows.}.
The deeper reason behind this has remained rather obscure, since these obviously have nothing
to do with supersymmetry.

Here we will elaborate on results obtained in \cite{Andrianopoli:2009je}, shewing
for the example of Einstein-Maxwell-(A)dS gravity in arbitrary dimension that the
first order flow equations satisfied by electrically charged static nonextremal black holes (found in part
for instance in \cite{Lu:2003iv,Miller:2006ay}) are just the 
expressions for the conjugate momenta in terms of derivatives of the principal function $W$ in a
Hamilton-Jacobi formalism \cite{deBoer:1999tgo}. Moreover, we will see that the expression for the potential in terms of a
`superpotential' is nothing else than the Hamilton-Jacobi equation for zero energy. The fact that a
nonextremal black hole solution arises from a first order system via a superpotential construction is thus
not surprising at all.

We will also find that (for the theory under consideration) there exist actually two different branches of 
solutions to the HJ equation. This leads to two distinct sets of flow equations, that share the same black
hole solutions.

Guided by the structure of $W$ in the Einstein-Maxwell-$\Lambda$ case, one can try to generalize
our analysis for instance to $N=2$ gauged supergravity in four dimensions, where the superpotential
in the BPS case is known both for $\text{U}(1)$ Fayet-Iliopoulos gauging \cite{DallAgata:2010ejj} and
for coupling to hypermultiplets, when abelian isometries of the quaternionic hyperscalar target manifold
are gauged \cite{Klemm:2016wng}. Unfortunately it turns out that the principal function $W$ for
nonextremal black holes is not straightforwardly generalizable to the matter-coupled case.
Nevertheless, we show (for the example of a particular prepotential) that there exist several conserved
charges that allow a partial separation of variables in the HJ equation. Among these conserved charges
there is the one originally introduced for ungauged supergravity in \cite{Trigiante:2012eb} and
subsequently adapted to the gauged theory in \cite{Klemm:2016wng}. Moreover, it was recently
found \cite{Cacciatori:2016xly} that $N=2$, $d=4$ $\text{U}(1)$ Fayet-Iliopoulos gauged supergravity
enjoys residual symmetries that essentially involve the stabilization of the symplectic vector of gauge 
couplings (FI parameters) under the action of the U-duality symmetry of the ungauged theory.
This provides additional conserved charges.

The remainder of this paper is organized as follows: In the next section, we consider
Einstein-Maxwell-(A)dS gravity in arbitrary dimension, adopt an ansatz for electrically charged static
black holes whose event horizon is a generic $(d-2)$-dimensional Einstein space, and determine
the one-dimensional effective action from which one can derive the equations of motion. In section
\ref{sec:int-HJ} we integrate the Hamilton-Jacobi equation associated to this mechanical system
in full generality and show that there are two branches of solutions. This leads to two different sets of
first order flow equations, one of which coincides with that found in \cite{Lu:2003iv}. In section
\ref{sec:sugra} we analyze how much of these integrability properties\footnote{Integrability in presence
of a cosmological constant was studied before in \cite{Charmousis:2006fx,Astorino:2012zm,
Leigh:2014dja,Klemm:2015uba}.} can be generalized to matter-coupled
$N=2$, $d=4$ gauged supergravity. We conclude in \ref{sec:final} with some final remarks.

\section{Static black holes in Einstein-Maxwell-(A)dS gravity}

We consider $d$-dimensional Einstein-Maxwell-(A)dS gravity, whose action is given by
\begin{equation}
S = \frac1{16\pi G_d}\int\D^d x\sqrt{-g}\left(R - F_{\mu\nu}F^{\mu\nu} - 2\Lambda\right)\,,
\label{eq:EML action}
\end{equation}   
with $d>3$. This is the simplest model that can be embedded (at least for some $d$) in $N=2$ gauged
supergravity. The equations of motion following from \eqref{eq:EML action} are
\begin{equation}
R_{\mu\nu} - \frac12 R g_{\mu\nu} + \Lambda g_{\mu\nu} = 2\left(F_{\mu\sigma}{F_{\nu}}^\sigma -
\frac14 g_{\mu\nu} F_{\sigma\rho}F^{\sigma\rho}\right)\,, \qquad
\nabla_{\mu}F^{\mu\nu} = 0\,, \label{eq:eom}
\end{equation}
where $F=\D A$.
For future convenience we report the trace and the traceless part of the Einstein equations that
respectively read
\begin{equation}
\begin{split}
& R - \frac{2 d}{d-2}\Lambda - \frac{d-4}{d-2} F^{\mu\nu} F_{\mu\nu} = 0\,, \\
& R_{\mu\nu} - 2 {F_{\mu}}^\sigma F_{\nu\sigma} - \frac2{d-2}\Lambda g_{\mu\nu} + \frac1{d-2} 
g_{\mu\nu} F^{\sigma\rho} F_{\sigma\rho} = 0\,. \label{eq:Einsplit}
\end{split}
\end{equation}

\subsection{Electrically charged black holes}

In what follows we shall consider electrically charged static black holes whose horizon is a
$(d-2)$-dimensional Einstein space\footnote{For $d>5$ this does not necessarily imply that the
horizon has constant curvature.}. The metric and the gauge field have the form 
\begin{equation}
\D s^2_d = - e^{-2(d-3)U}\D t^2 + e^{2U-2(d-4)\psi}\D r^2 + e^{2(U+\psi)}\D\Omega_{\kappa, d-2}^2\,,
\qquad A = A_t\D t\,, \label{eq:ans1}
\end{equation} 
where the functions $U$, $\psi$ and $A_t$  depend only on the coordinate $r$. The metric in
\eqref{eq:ans1} has the warped product structure
\begin{equation}
\D s^2_d = \tilde g_{ab}\D x^a\D x^b + f^2(x)\hat g_{ij}\D y^i\D y^j\,, \label{eq:warped}
\end{equation}
where the $(d-2)$-dimensional fiber with metric $\hat g_{ij}\D y^i\D y^j=\D\Omega_{\kappa, d-2}^2$ is
a generic Einstein space, i.e., $\hat R_{ij} = (d-3)\kappa\hat g_{ij}$. The nonvanishing components of the
Ricci tensor in $d$ dimensions are thus given by \cite{ONeill}
\begin{equation}
\begin{split}
& R_{ab} = \tilde{R}_{ab} - \frac{d_F}f\tilde\nabla_a\tilde\nabla_b f\,, \\
& R_{ij} = \hat{R}_{ij} - \hat g_{ij}\left(f\tilde\nabla_a\tilde\nabla^a f  + (d_F - 1)\tilde g^{ab}\partial_a f 
\partial_b f \right)\,, \label{eq:Ric-warped}
\end{split}
\end{equation}
where $d_F >1$ is the dimension of the fiber and $\tilde\nabla_a$ denotes the covariant derivative 
constructed with the Levi-Civita connection for $\tilde g_{ab}$.

\subsection{Effective action}

The Maxwell equations for the ansatz \eqref{eq:ans1} are solved by
\begin{equation}
F = - Q e^{-2(d-3)(U + \psi)}\D t\wedge\D r\,,
\end{equation}
where $Q$ is an integration constant corresponding to the electric charge. Using \eqref{eq:Ric-warped}
it is straightforward to shew that the Einstein equations \eqref{eq:Einsplit} boil down to three
ordinary differential equations that can be derived form the one-dimensional effective action
\begin{equation}
S_{\text{eff}} = \int\D r L = \int\D r\left(e^{2(d-3)\psi}(U'^2 - \psi'^2) - V_{\text{eff}}\right)\,, 
\label{eq:effsys}
\end{equation}
with the potential
\begin{equation}
V_{\text{eff}} = \kappa - \frac{2 Q^2}{(d-3)(d-2)} e^{-2(d-3)(U + \psi)} - \frac{2\Lambda}{(d-3)(d-2)}
e^{2(U + \psi)}\,, \label{eq:Veff}
\end{equation}
if we impose in addition the zero energy condition
\begin{equation}
e^{2(d-3)\psi}(U'^2 - \psi'^2) + V_{\text{eff}} = 0\,. \label{eq:H=0}
\end{equation}
To be concrete, the equation of motion for $U$ is proportional to the $tt$-component of \eqref{eq:Einsplit}, 
while the one for $\psi$ is a linear combination of the $tt$- and $rr$-components. Moreover,
from the first of \eqref{eq:Einsplit} and the $tt$-component one gets \eqref{eq:H=0}. The Einstein
equations along the fiber are automatically satisfied.\\
The conjugate momenta and Hamiltonian of the dynamical system \eqref{eq:effsys} are respectively
given by
\begin{equation}
\begin{split}
&p_U = \frac{\partial L}{\partial U'} = 2 e^{2(d-3)\psi} U'\,, \qquad p_{\psi} = \frac{\partial L}{\partial \psi'}
= - 2 e^{2(d-3)\psi}\psi'\,, \\
&H_{\text{eff}}(p_U, p_{\psi}, U, \psi) = \frac14 e^{-2(d-3)\psi}(p_U^2 - p_{\psi}^2) + V_{\text{eff}}\,.
\end{split}
\label{eq:Ham}
\end{equation}

\section{Integration of the Hamilton-Jacobi equation}
\label{sec:int-HJ}

The Hamilton-Jacobi equation associated to \eqref{eq:Ham} reads
\begin{equation}
H_{\text{eff}}(\partial_U S, \partial_{\psi} S, U, \psi) + \frac{\partial S}{\partial r} = 0\,. \label{eq: eqHJ}
\end{equation}
Since $H_{\text{eff}}$ does not depend explicitely on $r$ we set
\begin{equation}
S = 2 W(U,\psi) - E r\,,
\end{equation} 
such that \eqref{eq: eqHJ} reduces to
\begin{equation}
e^{-2(d-3)\psi}(W_U^2 - W_{\psi}^2) + V_{\text{eff}} = E\,, \label{eq:HJred}
\end{equation}
where $W_U$ and $W_{\psi}$ are respectively the partial derivatives of $W$ w.r.t.~$U$ and $\psi$. 
Inspired by \cite{Leigh:2014dja, Klemm:2015uba}, we define a new set of coordinates
\begin{equation}
X = e^{(d-3)(U + \psi)}\,, \qquad Y = e^{-2(d-3)U}\,,
\end{equation}
for which \eqref{eq:HJred} becomes
\begin{equation}
\frac{4(d-3)^2}{X^2}\left(Y W_Y^2 - X W_X W_Y\right) - \frac{2 Q^2}{(d-2)(d-3) X^2}  - \frac{2\Lambda
X^{\frac2{d-3}}}{(d-2)(d-3)} = \hat E\,, \label{eq: HJnew}
\end{equation} 
where $\hat E=E-\kappa$. To avoid loss of information $E$ will be set to zero, as required by
\eqref{eq:H=0}, only at the end of the integration procedure. The reason for this is that, in order to solve
the dynamics algebraically, one needs \eqref{eq:din1} and \eqref{eq:din2}, therefore we set $E=0$ only
after these equations have been obtained.

\subsection{First solution}

Applying the method of characteristics yields
\begin{equation}
\frac{\D W_Y}{W_Y} = \frac{\D X}X\,, \label{eq:ODE1}
\end{equation}
and thus $W_Y=aX$, where $a$ is an integration constant. The solution of this equation boils down
to $W(X,Y)= a Y X +\omega(X)$ that inserted into \eqref{eq: HJnew} leads to an ODE
\begin{equation}
-4 a (d-3)^2 \omega_X - \frac{2 Q^2}{(d-2)(d-3) X^2} - \frac{2\Lambda X^{\frac2{(d-3)}}}
{(d-2)(d-3)} = \hat E\,,
\end{equation} 
that can be easily integrated to give
\begin{equation}
S_1 = 2a Y X + \frac1{2 a (d-3)^2}\left(\frac{2 Q^2}{(d-2)(d-3) X} - \frac{2\Lambda X^{\frac{d-1}{d-3}}}
{(d-1)(d-2)} - \hat E X\right) - E r + C\,. \label{eq:gensol1}
\end{equation}
This contains three integration constants $C,E$ and $a$, where the latter must be different from zero.
Using
\begin{equation}
\frac{\partial S_1}{\partial E}\Big{|}_{E=0} = c_1\,, \qquad\frac{\partial S_1}{\partial a}\Big{|}_{E=0} = c_2\,,
\label{eq:din1}
\end{equation}
where $c_1$ and $c_2$ denote arbitrary constants, the dynamics can be solved algebraically, with the
result
\begin{equation}
\begin{split}
& X = -2a (d-3)^2 (r + c_1)\,, \\
& Y = \frac{c_2}{2 X} + \frac{Q^2}{2 a^2 (d-2)(d-3)^3 X^2} + \frac{\kappa}{4 a^2 (d-3)^2} -
\frac{\Lambda X^{\frac2{d-3}}}{2 a^2(d-1)(d-2)(d-3)^2}\,. \label{eq:sol1}
\end{split}
\end{equation}
In terms of $Y$ and the new radial coordinate $R=X^{\frac1{d-3}}$, the solution \eqref{eq:ans1} becomes
\begin{equation}
\begin{split}
\D s^2_d &= -Y\D t^2 + \frac{\D R^2}Y + R^2\D\Omega^2_{\kappa,d-2}\,, \qquad
F = \frac Q{R^{d-2}}\D t\wedge\D r\,, \\
Y &= \kappa -\frac{2 M}R + \frac{2 Q^2}{(d-2)(d-3) R^{2(d-3)}} - \frac{2\Lambda R^2}{(d-1)(d-2)}\,.
\label{eq:RN(A)dS}
\end{split}
\end{equation}
Here we fixed $a^2=\frac1{4(d-3)^2}$ (which can always be achieved by rescaling the coordinates
appropriately) and defined $c_2=-4M$. \eqref{eq:RN(A)dS} is the most general solution to the equations
of motion following from \eqref{eq:effsys}, and represents a generalization of the $d$-dimensional
Reissner-Nordstr\"om-(A)dS black hole to the case where the horizon is an arbitrary Einstein space.

In the original coordinates, Hamilton's characteristic function reads
\begin{displaymath}
W_1(U, \psi) = a e^{(d-3)(\psi - U)} + \frac{Q^2 e^{-(d-3)(U + \psi)}}{2 a (d-2)(d-3)^3} - \frac{\Lambda 
e^{(d-1)(U + \psi)}}{2 a (d-1)(d-2)(d-3)} + \frac{\kappa e^{(d-3)(U + \psi)}}{4 a (d-3)^2}\,.
\end{displaymath}
The expressions for the conjugate momenta
\begin{equation}
p_U = 2\frac{\partial W_1}{\partial U}\,, \qquad p_\psi = 2\frac{\partial W_1}{\partial\psi}\,,
\end{equation}
together with \eqref{eq:Ham}, lead to the first order flow equations
\begin{equation}
U' = e^{-2(d-3)\psi}\partial_U W_1(U, \psi)\,, \qquad \psi' = - e^{-2(d-3)\psi}\partial_{\psi} W_1(U, \psi)\,,
\label{eq:ffof}
\end{equation}
that are satisfied by the nonextremal black holes \eqref{eq:RN(A)dS}. Notice also that, using
\eqref{eq:ffof}, the action \eqref{eq:effsys} can be written as a sum of squares. This clarifies also
the reason for the very existence of first order equations for nonextremal black holes, namely
they are just the expressions for the conjugate momenta in terms of derivatives of the principal function
in a Hamilton-Jacobi formalism.

In the BPS case for $d=4$, one would expect to recover the supergravity BPS flow \cite{DallAgata:2010ejj}, in absence of vector multiplets, that is driven by\footnote{To derive \eqref{eq:Wbps} from the results of
\cite{DallAgata:2010ejj}, take the prepotential $F=-i(X^0)^2$ and a purely magnetic gauging with
FI-parameter proportional to $g$.}
\begin{equation}
W_{\text{BPS}}(U, \psi) = e^{-U} Q + e^{2\psi + U} g\,, \label{eq:Wbps}
\end{equation}
where $g$ is related to the cosmological constant by $\Lambda=-3g^2$. However, it is easy to see
that there is no limit in which \eqref{eq:Wbps} can arise from $W_1$. We shall come back to this
issue in the next subsection.

\subsection{Second solution}

Similar to what was done in \cite{Trigiante:2012eb} for $N=2$, $d=4$ ungauged supergravity, and
in \cite{Klemm:2016wng} for the abelian gauged case, we introduce the quantity
\begin{equation}
\mathbb Q \equiv e^{2(d-3)\psi}\frac{U' + \psi'}{d-3} + W\,.
\end{equation}
Using \eqref{eq:ffof} and the equations of motion following from the action \eqref{eq:effsys}, one
easily shows that $\mathbb Q'=0$, and thus $\mathbb Q$ is a constant of motion that can be
used to simplify \eqref{eq: HJnew}. In phase space we have
\begin{equation}
\mathbb Q = \frac{W_U - W_\psi}{d-3} + W = - 2YW_Y + W\,
\end{equation}
that implies $W(X,Y) = \mathbb Q + \sqrt{Y \omega(X)} $. 
Plugging this into \eqref{eq: HJnew} one gets the ODE
\begin{equation}
-4(d-3)^2\frac{\partial}{\partial X}\left(\frac{\omega}{X}\right) - \frac {2Q^2}{(d-2)(d-3) X^2}
-\frac{2\Lambda X^{\frac2{d-3}}}{(d-2)(d-3)} = \hat E\,.
\end{equation}
A final integration leads to the solution of the original differential
equation \eqref{eq: HJnew}\footnote{This solution was already found in \cite{Lu:2003iv} and for
$\kappa =0$ but with magnetic fluxes switched on in \cite{Lindgren:2015lia}.}  
\begin{equation}
S_2 = 2\mathbb Q - E r + 2\sqrt{-4A X Y + \frac{2 Q^2 Y }{(d-2)(d-3)^3} - \frac{\hat E X^2 Y }{(d-3)^2} -
\frac{2\Lambda X^{\frac{2d-4}{d-3}} Y}{(d-1)(d-2)(d-3)^2}}\,, \label{eq:S2}
\end{equation}
which has three arbitrary integration constants $\mathbb Q, E, A$, but in this case the parameter domain
is the whole $\mathbb R^3$. Using
\begin{equation}
\frac{\partial S_2}{\partial E}\Big{|}_{E=0} = c_3\,, \qquad \frac{\partial S_2}{\partial A}\Big{|}_{E=0} =  c_4\,,
\label{eq:din2}
\end{equation}
gives back \eqref{eq:sol1}, where
\begin{equation}
a = -\frac2{c_4}\,, \qquad c_1 = c_3\,, \qquad c_2 = -\frac{A c_4^2}2\,.
\end{equation}
To complete the comparison we evaluate 
\begin{displaymath}
\mathbb Q |_{W_1} = - a X Y + \frac{\kappa X}{4a(d-3)^2} + \frac{Q^2}{2a(d-2)(d-3)^2 X} -
\frac{\Lambda X^{\frac{d-1}{d-3}}}{2a(d-1)(d-2)(d-3)^2}\,.
\end{displaymath}
Plugging the solution \eqref{eq:sol1} into the rhs yields $2\mathbb Q=C-ac_2$.
In terms of $U$ and $\psi$, $W_2$ reads (setting $E=0$)
\begin{eqnarray}
\lefteqn{W_2(U,\psi) = \mathbb  Q} \nonumber \\
&+&\sqrt{A e^{(d-3)(\psi - U)} +\frac{2 Q^2 e^{-2(d-3)U}}{(d-2)(d-3)^3} + \frac{\kappa e^{2(d-3)\psi}}
{(d-3)^2} - \frac{2\Lambda e^{2(U + (d-2)\psi)}}{(d-1)(d-2)(d-3)^2}}\,, \label{eq:W_2Upsi}
\end{eqnarray}
which leads to the first order flow equations
\begin{equation}
U' = e^{-2(d-3)\psi}\partial_U W_2(U, \psi)\,, \qquad \psi' = - e^{-2(d-3)\psi}\partial_{\psi} W_2(U, \psi)\,.
\label{eq:sfof}
\end{equation}
\eqref{eq:sfof} and \eqref{eq:ffof} have different analytic forms, but share the same general class of
physical solutions. Notice also that, contrary to $W_1$, there is a well-defined limit in which
\eqref{eq:W_2Upsi} reduces to the BPS superpotential \eqref{eq:Wbps} for $d=4$, by setting
$A=0$, $\Lambda=-3g^2$ and imposing the Dirac-type quantization condition $2gQ=\kappa$.

The authors of \cite{Lu:2003iv} found that the potential \eqref{eq:Veff} can be expressed in terms
of a superpotential. One easily verifies that their superpotential (2.5) coincides with \eqref{eq:W_2Upsi}
and that eq.~(2.4) of \cite{Lu:2003iv} is just the Hamilton-Jacobi equation for zero energy.
The fact that a nonextremal black hole solution arises from a first order system via a superpotential
construction is thus not surprising at all.

\section{Matter-coupled $N=2$, $d=4$ gauged supergravity}
\label{sec:sugra}

In this section, we shall discuss possible generalizations of our formalism to $N=2$ supergravity
in four dimensions coupled to vector multiplets and with Fayet-Iliopoulos gauging. The analogue of the
one-dimensional effective action \eqref{eq:effsys} is then given by \cite{Klemm:2016wng}
\begin{equation}
S_{\text{eff}} = \int\D r\left(e^{2\psi}(U'^2 - \psi'^2 + g_{i\bar\jmath} z^{i\prime}
\bar z^{\bar\jmath\prime}) - V_{\text{eff}}\right)\,, \label{eq:Veff-vectors}
\end{equation}
with the potential
\begin{equation}
V_{\text{eff}} = \kappa - e^{-2(U+\psi)} V_{\text{BH}} - e^{2(U+\psi)} V_g(z,\bar z)\,, \label{eq:potsu}
\end{equation}
where \cite{DallAgata:2010ejj,Klemm:2016wng}
\begin{equation}
V_{\text{BH}} = g^{i\bar\jmath} D_i Z\bar D_{\bar\jmath}\bar Z + |Z|^2 = -\frac12 Q^T\ma{M} Q\,, \qquad
V_g = g^{i\bar\jmath} D_i{\cal L}\bar D_{\bar\jmath}\bar{\cal L} - 3 |{\cal L}|^2 \label{def-VBHVg}
\end{equation}
denote respectively the black hole- and scalar potential. In \eqref{def-VBHVg}, $D_i$ is the
K\"ahler-covariant derivative, $Z=\langle Q, {\cal V}\rangle$, ${\cal L}=\langle {\cal G}, {\cal V}\rangle$,
with the symplectic section $\cal V$ and the symplectic vectors of charges $Q$ and gauge couplings
${\cal G}$. $\ma M$ is the matrix defined in eq.~(2.7) of \cite{Klemm:2016wng}. Moreover
\begin{equation}
\langle A, B\rangle \equiv A^T\Omega B = A_\Lambda B^\Lambda - A^\Lambda B_\Lambda\,.
\end{equation}
Note that the target space of the one-dimensional sigma model \eqref{eq:Veff-vectors} is equipped with
the metric 
\begin{equation}
d\sigma^2 = e^{2\psi}(-d\psi^2 + dU^2 + g_{i\bar\jmath} dz^i d\bar z^{\bar\jmath})\,, \label{eq:mettar}
\end{equation}
and is thus a Lorentzian cone over a special K\"ahler manifold times a line, as can be seen by setting
$\tau=e^\psi$. The conjugate momenta and Hamiltonian read
\begin{equation}
\begin{split}
&p_U = 2 e^{2\psi} U'\,, \qquad p_\psi = -2 e^{2\psi}\psi'\,, \qquad p_i = e^{2\psi} g_{i\bar\jmath}
\bar z^{\bar\jmath\prime}\,, \qquad \bar p_{\bar\jmath} = e^{2\psi} g_{i\bar\jmath} z^{i\prime}\,, \\
&H_{\text{eff}} = e^{-2\psi}\left(\frac14 p_U^2 - \frac14 p_{\psi}^2 + g^{i\bar\jmath} p_i p_{\bar\jmath}
\right) + V_{\text{eff}}\,.
\end{split}
\end{equation}
If we set $S=2W-Er$, the reduced Hamilton-Jacobi equation becomes
\begin{equation}
e^{-2\psi}\left(W_U^2 - W_\psi^2 + 4g^{i\bar\jmath}\frac{\partial W}{\partial z^i}
\frac{\partial W}{\partial\bar z^{\bar\jmath}}\right) + V_{\text{eff}} = E\,.
\label{eq:eqHJ}
\end{equation}
As was shown for ungauged \cite{Trigiante:2012eb} and gauged supergravity \cite{Klemm:2016wng},
the quantity 
\begin{equation}
\mathbb Q \equiv e^{2\psi}(U' + \psi') + W\,,
\label{eq:consU4}
\end{equation}
is a first integral also in presence of the scalar fields $z^i$. $\mathbb Q$ is the Noether charge related
to the symmetry
\begin{equation}
\delta U = U_\epsilon -  U = \epsilon\,, \qquad \delta\psi = \psi_\epsilon - \psi = -\epsilon\,,
\label{scaling-symm}
\end{equation}
that leaves the potential \eqref{eq:potsu} and the action \eqref{eq:Veff-vectors} invariant (the latter up to
boundary terms). In fact, a function $W$, satisfying \eqref{eq:eqHJ} with $E=0$, drives a first order flow
\begin{equation}
U^{\prime} = e^{-2\psi} W_U\,, \qquad \psi^{\prime} = -e^{-2\psi} W_\psi\,, \qquad z^{i\prime} =
2 e^{-2\psi}g^{i\bar{\jmath}}\frac{\partial W}{\partial z^{\bar{\jmath}}}\,,
\end{equation}
and therefore the variation of \eqref{eq:Veff-vectors} for infinitesimal $\epsilon$ can be written as
\begin{equation}
\begin{split}
\delta S = S(U_\epsilon,\psi_\epsilon) - S(U,\psi) = & -2\epsilon\int\D r\left(e^{2\psi}(U'^2 - \psi'^2 +
4 g_{i\bar\jmath} z^{i\prime}\bar z^{\bar\jmath\prime})\right) \\
= & -2\epsilon\int\D r\left(U^{\prime} W_U + \psi^{\prime} W_\psi + z^{i\prime}\frac{\partial W}{\partial
z^i} + \bar z^{\bar{\jmath}\prime}\frac{\partial W}{\partial\bar z^{\bar{\jmath}}}\right) \\
= & -2\epsilon\int\D r\frac{d W}{dr}\,,
\end{split}
\end{equation}
which vanishes if we choose appropriate boundary conditions. Note that the transformation
\eqref{scaling-symm} is generated by the vector field
$\partial_U-\partial_\psi=\partial_U-\tau\partial_\tau$, which is a conformal Killing vector of the
Lorentzian cone \eqref{eq:mettar}. The fact that $\mathbb Q$ is the Noether charge related to
\eqref{scaling-symm} follows also from the inverse Noether theorem\footnote{See \cite{Banados:2016zim}
for a nice review.}: If $\mathbb Q$ is a conserved charge, then the transformation
\begin{equation}
\delta q^I = [q^I, \epsilon\mathbb Q] = \epsilon\frac{\partial\mathbb Q}{\partial p_I}\,, \qquad
\delta p_I = [p_I, \epsilon\mathbb Q] = -\epsilon\frac{\partial\mathbb Q}{\partial q^I}\,,
\end{equation}
where $[\,,\,]$ denotes the Poisson bracket, is a symmetry of the action.

As before, we introduce the coordinates
\begin{equation}
X = e^{U +\psi}\,, \qquad Y = e^{-2U}\,.
\end{equation}
Then the first integral \eqref{eq:consU4} becomes
\begin{equation}
\mathbb Q = - 2 Y W_Y + W\,, \label{eq:cons4}
\end{equation}
which can be easily integrated to give
\begin{equation}
W(X,Y,z,\bar z) = \mathbb Q + \sqrt{Y\omega(X,z,\bar z)}\,, \label{eq:def-omega}
\end{equation}
where $\omega$ is an integration `constant'. Using \eqref{eq:def-omega}, the Hamilton-Jacobi
equation \eqref{eq:eqHJ} boils down to
\begin{equation}
-\partial_X\frac{\omega}{X} + \frac1{\omega X^2} g^{i\bar\jmath}\frac{\partial\omega}{\partial z^i} 
\frac{\partial\omega}{\partial\bar{z}^{\bar\jmath}} - X^2 V_g -\frac1{X^2}V_{\text{BH}} + \kappa = E\,.
\label{eq:omegahj}
\end{equation}
A particular solution to \eqref{eq:omegahj} is the one found in \cite{DallAgata:2010ejj} by squaring the
action for the BPS case,
\begin{equation}
\omega_{\text{BPS}} = (Z - i X^2\ma L)(\bar{Z} + i X^2\bar{\ma L}) = |Z|^2 + X^4 |\ma L|^2
- i X^2 (\ma L\bar{Z} - \bar{\ma L} Z)\,. \label{eq:bpssol}
\end{equation}
Imposing $E=0$, as required by Einstein's equations, and using
\begin{equation}
\frac{\partial\omega}{\partial z^i} = (\bar{Z} + i X^2\bar{\ma L})(D_i Z - i X^2 D_i\ma L)\,,
\end{equation}
as well as the special K\"ahler geometry identity
\begin{equation}
\frac12 (\ma M - i\Omega) = \Omega\bar{\ma V}\ma V\Omega + \Omega D_i\ma V g^{i\bar{\jmath}} 
D_{\bar{\jmath}}\bar{\ma V}\Omega\,,
\end{equation}
it is only matter of some algebra to shew that \eqref{eq:bpssol} solves \eqref{eq:omegahj} if one imposes
the Dirac charge quantization condition
\begin{equation}
\langle\ma G, \ma Q\rangle = -\kappa\,.
\end{equation}
In the following subsection we shall consider a particular prepotential, for which the effective action 
\eqref{eq:Veff-vectors} has additional symmetries, that allow a further reduction of the Hamilton-Jacobi
equation \eqref{eq:omegahj}.

\subsection{Prepotential $F=-i X^0X^1$}

This simple model has only one complex scalar field $z$ parametrizing the Poincaré half-plane,
with K\"ahler metric
\begin{equation}
\D s^2 = \frac{\D z\D\bar{z}}{(z + \bar z)^2}\,,
\end{equation}
which has the three Killing vectors
\begin{equation}
v_1= i(\partial_z - \partial_{\bar z})\,, \qquad v_2 = z \partial_z + \bar z\partial_{\bar z}\,, \qquad
v_3 = \frac i2(\bar z^2\partial_{\bar z} - z^2\partial_z)\,.
\end{equation}
These are all symmetries of the ungauged theory, but in presence of a potential for the scalars only a
linear combination of them survives, as was shown in \cite{Cacciatori:2016xly} using the symplectic 
representation.

If we consider a configuration with only magnetic charges and purely electric gaugings, the HJ equation
\eqref{eq:omegahj} becomes for this prepotential
\begin{equation}
-\partial_X\frac{\omega}X + \frac1{\omega X^2} g^{z\bar z}\frac{\partial\omega}{\partial z}
\frac{\partial\omega}{\partial\bar{z}} + X^2\frac{g_0^2 + g_1^2 z\bar z + 2 g_0 g_1(z +
\bar z)}{z + \bar z} - \frac1{X^2}\frac{{p^1}^2 + {p^0}^2 z\bar z}{z + \bar z} + \kappa = E\,.
\label{eq:omegahj-F}
\end{equation}
The linear combination
\begin{equation}
v = \frac{g_0^2}{2 g_1^2} v_1 + v_3 = \frac i2\left(\frac{g_0^2}{g_1^2} - z^2\right)\partial_z -
\frac i2\left(\frac{g_0^2}{g_1^2} - \bar{z}^2\right)\partial_{\bar{z}}
\label{eq:sym}
\end{equation}
generates a symmetry of \eqref{eq:Veff-vectors} if one imposes the BPS condition \cite{Cacciatori:2009iz}
$p^0 g_0 = p^1 g_1$. It is straightforward to verify that this implies the existence of a further
conserved charge
\begin{equation}
\mathbb C = \frac i2\left(\frac{g_0^2}{g_1^2} - z^2\right)\frac{\partial\omega}{\partial z} - \frac i2 
\left(\frac{g_0^2}{g_1^2} - \bar{z}^2\right)\frac{\partial\omega}{\partial \bar z}\,.
\label{eq:consch}
\end{equation}
By introducing the new variables
\begin{equation}
z = \frac{g_0}{g_1}\tanh\left(\frac{g_0}{g_1}(u + i v)\right)\,, \qquad\bar{z} = \frac{g_0}{g_1} 
\tanh\left(\frac{g_0}{g_1} (u - i v)\right)\,,
\label{eq:coor}
\end{equation}
\eqref{eq:consch} can easily be integrated, with the result $\omega=2\mathbb C v +\alpha(u, X)$.
Plugging this into \eqref{eq:omegahj-F}, the HJ equation assumes the form
\begin{equation}
-\partial_X\frac{\alpha}X + \left(\frac{g_1}{g_0}\right)^2\frac{\sinh^2(2 g_0 u/g_1)}{4X^2}
\frac{\alpha_u^2 + 4\mathbb C^2}{\alpha + 2\mathbb C v} - X^2 V_g(u) - \frac1{X^2} V_{\text{BH}}(u)
+ \kappa = E\,, \label{eq:alphahj}
\end{equation}
where 
\begin{equation}
V_{\text{BH}}(u) = \frac{(p^0)^2 g_0}{g_1\tanh(2 g_0 u/g_1)}\,, \qquad V_{g}(u) =
-\frac{g_0 g_1}{\tanh(2 g_0 u/g_1)} - 2 g_0 g_1\,.
\end{equation}
It is easy to see that \eqref{eq:alphahj} can be satisfied for all $v$ only if $\mathbb C = 0$\footnote{This
sort of `axion-free' condition is probably related to the special choice of purely electric gaugings and
only magnetic charges, so we don't expect that $\mathbb C$ vanishes in a more general setting.},
so that we have
\begin{equation}
-\partial_X\frac{\alpha}X + \left(\frac{g_1}{g_0}\right)^2\frac{\sinh^2(2 g_0 u/g_1)}{4X^2}
\frac{\alpha_u^2}{\alpha} - X^2 V_g(u) - \frac1{X^2} V_{\text{BH}}(u) + \kappa = E\,.
\label{eq:alphahjczero}
\end{equation}
For the prepotential under consideration, the BPS solution \eqref{eq:bpssol} reads
\begin{equation}
\omega_{\text{BPS}}(X, z, \bar z) = \frac{(p^1 + p^0 z - X^2 (g_0 + g_1 z))(p^1 + p^0\bar z - X^2
(g_0 + g_1\bar z))}{2(z +\bar z)}\,.
\end{equation}
Imposing $g_0 p^0 = g_1 p^1$ and using the coordinates \eqref{eq:coor}, this leads to
\begin{equation}
\alpha_{\text{BPS}}(X, u) = \frac{p^1(p^0 - g_1 X^2)^2 e^{4 p^1 u/p^0}}{p^0 (e^{4 p^1 u/p^0} - 1)}\,.
\label{eq:alpha-BPS}
\end{equation}
It is interesting to note that the variables $X$ and $u$ separate in \eqref{eq:alpha-BPS}. This suggests
to use a product ansatz $\alpha(X,u)=\xi(X)\mu(u)$ in order to get something more general than
\eqref{eq:alpha-BPS}. Unfortunately, plugging this into \eqref{eq:alphahjczero} gives back precisely 
\eqref{eq:alpha-BPS}. Another possibility is inspired by the comparison with \eqref{eq:S2}
(for $d=4$), which contains, in addition to quartic, quadratic and $X$-independent terms
that appear also in \eqref{eq:alpha-BPS}, a linear piece in $X$ proportional to the constant $A$ that
is essentially a nonextremality parameter (or black hole mass). One may thus try
\begin{equation}
\alpha(X,u) = \sum_{n=0}^4\alpha_n(u) X^n\,,
\end{equation}
where (to be still more general) we added a cubic term as well. However, one can check that, using this
ansatz in  \eqref{eq:alphahjczero} leads to an overdetermined system that admits a solution only for 
$\alpha_1=\alpha_3=0$, namely \eqref{eq:alpha-BPS}.

It remains to be seen if there exist additional conserved charges associated to hidden symmetries
of the action \eqref{eq:Veff-vectors}, that would allow to completely separate the Hamilton-Jacobi
equation \eqref{eq:eqHJ}. Note in this context that the transformation \eqref{scaling-symm} acts only
on $U$ and $\psi$ but not on the scalars $z^i$, whereas \eqref{eq:sym} touches only the $z^i$
but not the metric components $U$ and $\psi$. There might thus exist (at least for some specific
models) more complicated symmetry transformations involving all the dynamical variables.
We hope to come back to a systematic analysis of this issue in a future publication.

\section{Final remarks}
\label{sec:final}

In this paper we considered electrically charged static nonextremal black holes in $d$-dim\-ens\-ional
Einstein-Maxwell-(A)dS gravity, whose horizon is a generic Einstein space in $d-2$ dimensions. We have
shown that for this system the Hamilton-Jacobi equation is exactly integrable and admits two branches
of solutions.
One of them exhibits a non-simply connected domain of integration constants and does not reduce to
the well-known solution for the $d=4$ BPS case. The principal functions generate two first order flows
that are analytically different, but support the same general solution. One of the two sets of flow
equations corresponds to those found in \cite{Lu:2003iv} and (for 
$d=4$ and $\Lambda=0$) in \cite{Miller:2006ay}. We clarified thus also the
reason for the very existence of first order equations for nonextremal black holes, namely, they are just
the expressions for the conjugate momenta in terms of derivatives of the principal function in
a Hamilton-Jacobi formalism.

In the last part of our paper we also analyzed if these integrability properties continue to hold for
matter-coupled $N=2$, $d=4$ gauged supergravity. Unfortunately it turned out that the principal
function $W$ for nonextremal black holes is not straightforwardly generalizable to this case.
Still, we showed (for the example of a particular model) that there exist several conserved
charges that allow a partial separation of variables in the HJ equation.
These conserved charges comprise
the one originally introduced for ungauged supergravity in \cite{Trigiante:2012eb} and
subsequently adapted to the gauged theory in \cite{Klemm:2016wng}, as well as those associated to
the symmetries recently discovered in \cite{Cacciatori:2016xly}. We pointed out the possible existence
of additional hidden symmetries of the one-dimensional effective action \eqref{eq:Veff-vectors}
that involve simultaneous transformations of the dynamical variables of both the metric and the
scalar sector.

One might ask if there exist covariantly constant spinors related to the first order equations. The authors
of \cite{Miller:2006ay} have shown that the nonextremal Reissner-Nordstr\"om solution cannot admit
(generalized) Killing spinors in 3+1 dimensions, but it is supersymmetric in a lower-dimensional
effective theory. It might be, however, that the nonextremal black holes considered in this paper
possess so-called conformal Killing spinors (CKS, cf.~e.g.~\cite{Baum:2002jt} for a review of this topic).
Note in this context that both the (nonextremal) Kerr metric and all other type II-II vacuum spacetimes
do admit a CKS \cite{Walker:1970un}. We hope to come back to this point in a future publication.

\section*{Acknowledgements}

This work was supported partly by INFN.

\end{document}